\documentclass{aastex}
\usepackage{spr-astr-addons}
\usepackage{url}

\RequirePackage{color}

\begin{document}
\def\be{\begin{equation}}
\def\ee{\end{equation}}
\def\bea{\begin{eqnarray}}
\def\eea{\end{eqnarray}}
\def\c{\cite}
\def\nn{\nonumber}
\def\mcr{{{\rm M_{cr}}}}
\def\xo{{X_{0}}}
\def\dm{\Delta {\rm M}}
\def\ms{\ifmmode {\rm M}_{\odot}{ ~ } \else ${\rm M}_{\odot}{ ~ }$ { }\fi}
\def\mb{m_{B}}
\def\bo{B_{o}}
\def\cxo{C(x_{o})}
\def\rsix{R_{6}}
\def\rnin{R_{9}}
\def\vr{v_{{\rm r}}}
\def\vro{v_{{\rm ro}}}
\def\vt{v_{\theta}}

\def\et{ {\it et al.}}
\def\la{ \langle}
\def\ra{ \rangle}
\def\ov{ \over}
\def\ep{ \epsilon}

\def\ep{\epsilon}
\def\th{\theta}
\def\ga{\gamma}
\def\Ga{\Gamma}
\def\la{\lambda}
\def\si{\sigma}
\def\al{\alpha}
\def\pa{\partial}
\def\de{\delta}
\def\De{\Delta}
\def\rsr{{r_{s}\over r}}
\def\rmo{{\rm R_{M0}}}
\def\rrm{{R_{{\rm M}}}}
\def\rra{{R_{{\rm A}}}}

\def\mdot{\ifmmode \dot M \else $\dot M$\fi}    % accretion rate
\def\mxd{\ifmmode \dot {M}_{x} \else $\dot {M}_{x}$\fi}
\def\med{\ifmmode \dot {M}_{Edd} \else $\dot {M}_{Edd}$\fi}
\def\bff{\ifmmode B_{{\rm f}} \else $B_{{\rm f}}$\fi}

\def\apj{\ifmmode ApJ \else ApJ \fi}    % lower
\def\apjl{\ifmmode  ApJ \else ApJ \fi}    %
\def\apjs{\ifmmode  ApJS \else ApJS \fi}
\def\aap{\ifmmode A\&A \else A\&A\fi}
\def\aaps{\ifmmode A\&AS \else A\&AS\fi}    %
\def\mnras{\ifmmode MNRAS \else MNRAS \fi}    %
\def\nat{\ifmmode Nature \else Nature \fi}
\def\prl{\ifmmode Phys. Rev. Lett. \else Phys. Rev. Lett.\fi}
\def\prd{\ifmmode Phys. Rev. D. \else Phys. Rev. D.\fi}
\def\pasp{\ifmmode  PASP \else PASP \fi}

\def\ms{\ifmmode {\rm M}_{\odot} \else ${\rm M}_{\odot}$\fi}    % lower
\def\na{\ifmmode \nu_{A} \else $\nu_{A}$\fi}    % Alfven frequency
\def\nk{\ifmmode \nu_{K} \else $\nu_{K}$\fi}    % Keplerian frequency
\def\ns{\ifmmode \nu_{{\rm s}} \else $\nu_{{\rm s}}$\fi}
\def\no{\ifmmode \nu_{1} \else $\nu_{1}$\fi}    % lower
\def\nt{\ifmmode \nu_{2} \else $\nu_{2}$\fi}    % upper
\def\ntk{\ifmmode \nu_{2k} \else $\nu_{2k}$\fi}    % upper
\def\dnmax{\ifmmode \Delta \nu_{max} \else $\Delta \nu_{2max}$\fi}
\def\ntmax{\ifmmode \nu_{2max} \else $\nu_{2max}$\fi}    % upper
\def\nomax{\ifmmode \nu_{1max} \else $\nu_{1max}$\fi}    % upper
\def\nh{\ifmmode \nu_{\rm HBO} \else $\nu_{\rm HBO}$\fi}    % HBO
\def\nqpo{\ifmmode \nu_{QPO} \else $\nu_{QPO}$\fi}    % HBO
\def\nz{\ifmmode \nu_{o} \else $\nu_{o}$\fi}    % HBO
\def\nht{\ifmmode \nu_{H2} \else $\nu_{H2}$\fi}    % HBO
\def\ns{\ifmmode \nu_{s} \else $\nu_{s}$\fi}    % stellar
\def\nb{\ifmmode \nu_{{\rm burst}} \else $\nu_{{\rm burst}}$\fi}
\def\nkm{\ifmmode \nu_{km} \else $\nu_{km}$\fi}    % stellar
\def\ka{\ifmmode \kappa \else \kappa\fi}    % stellar
\def\dn{\ifmmode \Delta\nu \else \Delta\nu\fi}
\def\dm{\ifmmode \Delta{}M \else \Delta{}M\fi}
\def\mdotsix{\ifmmode\dot{M}_{16} \else \dot{M}_{16}\fi}
\def\ps{\ifmmode P_{spin} \else P_{spin} \fi}

\def\sax{\ifmmode SAX J 1808.4-3658 \else SAX J 1808.4-3658\fi}

 \def\pspin{\ifmmode P_{s} \else $P_{s}$\fi}

\renewcommand{\vec}[1]{\mbox{\boldmath $\displaystyle #1$}}
\newcommand{\grad}{\vec{\nabla}}
\newcommand{\lap}{\nabla^2}
\newcommand{\vdot}{\vec{\cdot}}
\newcommand{\vcross}{\vec{\times}}
\newcommand{\divr}{\grad\vdot\,}
\newcommand{\curl}{\grad\vcross\,}
\newcommand{\avZ}{\left<Z\right>}
\def\rhof{\ifmmode \rho_{5} \else \rho_{5}\fi}
\def\rhos{\ifmmode \rho_{6} \else \rho_{6}\fi}
\def\mdotcr{\ifmmode \dot{M}_{cr} \else  \dot{M}_{cr}\fi}
\def\tohm{\ifmmode t_{ohmic} \else  t_{ohmic} \fi}

\def\tdif{\ifmmode t_{diff} \else  t_{diff} \fi}
\def\tacc{\ifmmode t_{accr} \else  t_{accr} \fi}
\newcommand{\gsimeq}{\mbox{$\, \stackrel{\scriptstyle >}{\scriptstyle
\sim}\,$}}
\newcommand{\lsimeq}{\mbox{$\, \stackrel{\scriptstyle <}{\scriptstyle
\sim}\,$}}

\title{%Study of
Binary Pulsars in Magnetic Field versus Spin Period Diagram}
\slugcomment{Not to appear in Nonlearned J., 45.}
%% Running heads
\shorttitle{Short article title}
\shortauthors{Autors et al.}

\author{Y. Y. Pan\altaffilmark{1,2}} \and \author{N. Wang\altaffilmark{1}}
\affil{Xinjiang  Astronomical Observatories, Chinese Academy of Sciences,
Xinjiang 830011, China. {\bf panyuanyue@xao.ac.cn}}
\and
\author{C. M. Zhang\altaffilmark{2}}
\affil{National Astronomical Observatories, Chinese Academy of Sciences,
Beijing, 100012, China. {\bf zhangcm@bao.ac.cn}}
\email{panyuanyue@xao.ac.cn}

%\altaffiltext{1}{First Alternate Affilation.}
%\altaffiltext{2}{Second Alternate Affilation.}
%\altaffiltext{3}{Third Alternate Affilation.}

\begin{abstract}
We analyzed 186 binary pulsars (BPSRs) in the magnetic field versus
spin period (B-P) diagram, where their relations to the millisecond pulsars (MSPs)
can be clearly shown. Generally, both BPSRs and MSPs are believed to be recycled
and spun-up in binary accreting phases, and evolved below the spin-up line
setting by the Eddington accretion rate ($\dot{M}{\bf \simeq}10^{18} g/s$).
It is noticed that most BPSRs are distributed around the spin-up line with mass
accretion rate $\dot{M}=10^{16} g/s$ and almost all MSP samples lie above the
spin-up line with $\dot{M}\sim10^{15} g/s$. Thus, we calculate that a minimum
accretion rate ($\dot{M}\sim 10^{15} g/s$) is required for the MSP formation, and
physical reasons for this are proposed.
In the B-P diagram, the positions of BPSRs and their relations to the binary
parameters, such as the companion mass, orbital period and eccentricity, are
illustrated and discussed.
In addition, for the seven BPSRs located above the limit spin-up line, possible
causes are suggested.
\end{abstract}


%\section*{}
%\label{sec:intro}

\section{Introduction}

The pulsars (PSRs) in the binary systems are always convenient for us to determine their parameters such as the mass, origin and evolution.
The majority of the binary pulsars (BPSRs) are millisecond pulsars (MSPs) with the spin period less than 20 milliseconds \citep{sta04}.
Based on the data from the ATNF Pulsar Catalogue in November 2012, it has been found 2008
PSRs including 221 MSPs (136 binary PSRs versus 85 isolated ones). The companions of 186
binary PSRs are white dwarfs (170), neutron stars (9), main sequence stars (4) and planets (3), respectively.
Much progress has been achieved in understanding the formation and evolution of the
binary pulsars (BPSRs) \citep{bha95, sta04, man04, lor08, tau12s, tau12}.

The distributions of the magnetic field ($B$) and spin period ($P$) are both almost
bimodal with a dichotomy between the normal pulsars (PSRs) ($B\sim10^{11-13} G$,
$P\sim 0.1 s - 10 s$) formed directly by the supernova explosions and MSPs
($B\sim10^{8-9} G$, $P\sim{1.4 ms-20 ms}$) that are recycled in the binary systems
\citep{alp82, rad82, lor08, kas10, wang11}.
In the binary systems, with the accreting matter of $\sim 0.1-0.2\ms$ from their companions,
the neutron stars (NSs) can be spun-up to several milliseconds, while their magnetic fields
decay to $\sim 10^{8-9} G$ \citep{bha91, bha95, heu95, mel05, zhang06, wang11}.
%\citep{bha91, bha95, heu95, mel05, zhang06, wang11}.
%
With the 3D simulations of magnetohydrodynamics (MHD), the magnetic field affected by the
accretion flow is depicted clearly \citep{kul08}. When the accretion phase is finished,
the radio emissions of the fast rotating NSs can be detected as MSPs \citep{alp82, rad82, tau12}.
%\citep{alpar82, rad82, tau12.}
%

Evaporation of the companion star by the pulsar radiation explain explains why
some MSPs are single \citep{klu88,  hes08}.
%
%Sometimes a single MSP can be formed because of the evaporation of companion by the MSP
%radiation \citep{klu88,  hes08}, which is ascribed to the fact that
%some of the MSPs have no companions.
This is more complicated in globular clusters:
a companion can be disrupted by encounter with another star.
%
%Direct evidence supporting the MSP spin-up model has been found, for instance, from the
%accreting X-ray MSP SAX J1808.4-3658 with $B\sim10^{8} G$ and $P=2.49 ms$ is discovered
%in low mass X-ray binary system \citep{wij98}, and the double pulsar PSR 0737-3039A/B,
%which consists of one MSP and a normal PSR \citep{bur03, heu04, lyn04}.
%
Direct evidence supporting the MSP spin-up model has been found, for instance, the
accreting X-ray MSP SAX J1808.4-3658 ($B\sim10^{8} G$, $P=2.49 ms$)
discovered in the low mass X-ray binary system \citep{wij98}, and the double
pulsar PSR 0737-3039A/B (a MSP and a normal PSR in the binary system) \citep{bur03, heu04, lyn04}.

 BPSRs are the most accurate clocks in the unverse that can be exploited to form
a MSP detect gravitational waves \citep{man04, lor08,hob10}.
BPSRs are the most accurate clocks in the unverse
%\citep{man04, lor08}.
%that can be exploited to form a MSP array to detect gravitational waves \citep{hob10}.
%\citep{hob10}.
%
In addition, the BPSRs are significant systems for testing general relativity effects,
e.g., the gravitational-radiation-induced orbit shrinking, gravitational red-shift and
Shapiro delay \citep{kra06}, based on which the NS mass can be measured in double
neutron star systems using the periastron advance and with the high precision \citep{zhang11}
%\citep{zhang11}.
%

This paper mainly focuses on the analysis  of 186 BPSRs in the B-P diagram with the
data from ATNF  pulsar catalogue (Manchester et al. 2005).%\citep{man05}.
To understand the properties of these pulsars, we pay attention to their B-P
positions relative to the spin-up lines, in relation to the companion mass,
eccentricity and orbit period.

The  structure of the paper is  organized as follows: Sec.2 presents the detail
study of the spin up lines under the different accretion rates. Here we also give
the implications of the minimum accretion rate on the millisecond pulsar formation.
In Sec.3, the distribution  characteristics of BPSPs in the B-P diagram as a function
of the binary parameters and the evolutionary scenario of binary pulsar are discussed.
In Sec.4, we suggest the causes why seven BPSRs lie  above the Eddington limited
spin-up line. The discussions and conclusions are given in Sec.5.

\section{Pulsars in B-P diagram and the evolutionary implications}

\subsection{The Spin-up line}
%\subsection{The equilibrium spin period}

%Assuming that the magnetic field of a NS has a symmetry axis aligned with
%the rotation axis, when the NS spins up to the Kepler orbital period at the
%magnetosphere radius, a spin-up line will be get, which is also called the
%equilibrium spin period line expressed the relationship between the magnetic
%field and the spin period
If a NS, which has a symmetry axis of the magnetic field aligned with the rotation axis,
spins up to the Kepler orbital period at the magnetosphere radius, a spin-up line will
be obtained. It is also called the equilibrium spin period line expressed the
relationship between the magnetic field and the spin period
\citep{gho77, sha83, bha91, gho92, bur02, fra02, cam07}.
However, the magnetosphere radius depends on
many factors, such as the structure of the accretion disk and mass transfer form, etc
\citep{tau11, tau12}.%\citep{frank02, gho92, sha83}.

The equation of the spin-up line can be written as \citep{bha91}:
%\citep{bha91}:

\bea
\label{P}
P_{eq}=2.4 (ms)
B_9^{\frac{6}{7}}(\frac{M}{M_{\odot}})^{-\frac{5}{7}}(\frac{\dot{M}}{\dot{M}{_{Edd}}})^{-\frac{3}{7}}R_6^{\frac{18}{7}},
%&=& 1.9(ms)B_9^{6/7}(M/1.4M_{\odot})^{-5/7}(\dot{M}/\dot{M}{_{Edd}})^{-3/7}R_6^{16/7}\nonumber,
%=2.4 (ms)\cdot B_9^{6/7}\cdot (1.4)^{-5/7}
\eea
where $P_{eq}$ is the "equilibrium" spin period, $B_9$ is the dipole magnetic field in units
of $10^9G$, $\dot{M}$ is the accretion rate, $\dot{M}_{Edd}$ is the
"Eddington-limit" accretion rate ($10^{18} g/s$), $R_6$ is the stellar radius in units
of $10^6$cm and $M$ is the NS mass.

The spin period ($P$) and the spin-down rate ($\dot{P}$) of a PSR can be combined to
yield an estimate of the strength of the dipole   magnetic field \citep{man77, lyn06}:
\bea
\label{B}
B&=&(\frac{3c^3IP\dot{P}}{8\pi^2R^6})^{\frac{1}{2}}\\
\nonumber
&=&3.2\times10^{19}({I\ov 10^{45} g cm^2})^{\frac{1}{2}}R_6^{-3}(P\dot{P})^{\frac{1}{2}},
\eea
where $I$ is the moment of inertial and $R$ is the radius of the PSR.
The moment of inertia is $I$ = $(2/5)MR^2$. To evaluate the
influence on the divided magnetic field by
the choice of NS mass and radius, we rewrite Eq.(\ref{B}) in terms of M and R:
\begin{equation}
%\begin{array}{l}
\label{Bsim}
 %(3C^3M/20\pi^2R^4)^{1/2}(P\dot{P})^{1/2}\\
B=3.4\times10^{19}(\frac{M}{1.4M_{\odot}})^{\frac{1}{2}}R_6^{-2}(P\dot{P})^{\frac{1}{2}}.
%\end{array}
\end{equation}
Eqs. (\ref{P}-\ref{Bsim}) show us that the NS  mass ($M$), radius ($R$) and accretion
rate ($\dot{M}$) influence on the position of the spin-up line in the B-P diagram,
and both NS masses and radius  have influences on the derived strength of the dipole magnetic
field. Mostly one takes a standard pulsar parameter ($M=1.4M_{\odot}$, $R=10^6 cm$)
in the three equations. However, some measured NS masses are
larger or smaller than $1.4M_{\odot}$. In binary systems, a mean value of $M=1.46\pm0.30M_{\odot}$
has been derived for the 61 measured NS masses \citep{zhang11}.
%\citep{zhang11}.
The NS radius has not yet been measured as accurate as its mass, and is usually estimated
in the range $10 \sim 20 km$ by means of the models \citep{mil98, mil02, lat04, hes06, hae07, zhang07}.
%\citep{mil98, mil02, lat04, hes06, hae07, zhang07}.
%

The spin-up line equation Eq.(1) and  magnetic field equation Eq.(3)
show the relationships  between the PSR magnetic field and NS radius
as $B\propto R^{-3}$ and $B\propto R^{-2}$, respectively. If a NS
radius changes from 10 km to 20 km, then the spin-up line will shift
down by Log(8)=0.9, whereas the PSR position will shift down by
Log(4)=0.6,  in Log(B) axis of the B-P diagram. That is to say, the
spin-up line will shift down  $\sim0.3$  relative to  the PSR
position in the Log(B) axis of the B-P diagram.

\begin{figure}
\includegraphics[width=7cm]{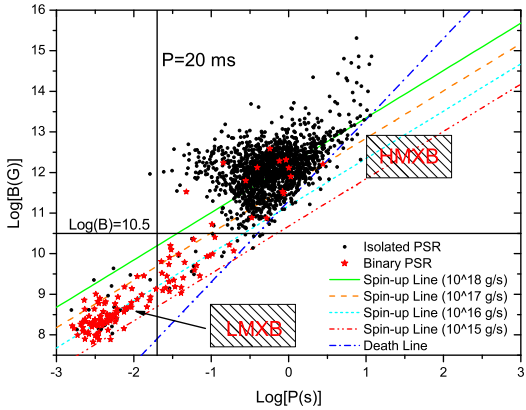}
\caption{The magnetic-field versus spin-period diagram for 2008 PSRs (data from ATNF pulsar
catalogue as at November 2012). 221 MSPs ($P<20 ms$) are shown,
including the isolated ones (85) and those in the binary systems (136). The solid, dash, short dash and dash dot-dot
lines stand for the spin-up lines with the different accretion rates $10^{18} g/s$, $10^{17} g/s$,
$10^{16} g/s$  and $10^{15} g/s$, respectively, as shown in Eq. (1). The dash dot line is the
death line by the theoretical model of deceasing the radio emission of pulsar (Bhattacharya \& van
den Heuvel 1991; Ruderman \& Sutherland 1975).
%\citep{bha91, rud75}.
The isolated and BPSRs are labeled by dots and stars, respectively.}\label{6}
\end{figure}

Compared with the affection of the radius  on the spin-up line and
PSR position in B-P diagram, the variation of accretion
rate  will give much more influence on the spin-up line.
From the observations, the accretion rates of the X-ray NSs span three
orders of magnitudes, e.g., from $10^{15} g/s$ to $10^{18} g/s$ \citep{has89, kli00, lam05},
and these variations will give rise to a considerately changes in the positions of the
spin-up line in the B-P diagram, which are shown in Fig.\ref{6} for accretion rates:
$10^{15} g/s$, $10^{16} g/s$, $10^{17} g/s$ and $10^{18} g/s$.

The distribution of the isolated  and binary PSRs are also shown in Fig.\ref{6}.
%It can be seen that more than half of the isolated PSRs are
% above the limit spin-up line ($10^{18} g/s$).
%
The majority of the MSPs are distributed around the spin-up line with $\mdot=10^{16} g/s$.
It seems likely that most MSPs may be born at the positions near the spin-up line with
$\mdot=10^{17} g/s$, which is the generally observed  accretion rate for the X-ray NSs
in the Low-Mass X-ray Binaries \citep{bra83, has89, kli00, liu07}.
%(Bradt \& McClintock 1983; Hasinger \& van der Klis 1989; van
%der Klis 2000; Liu, Paradijs \& van den Heuvel 2007).
%\citep{brad83, has89, kli00, liu07}.
%
This fact was also pointed out a decade ago by Lundgren et al. (1996) and Hansen and
Phinney (1998).

\subsection{Implications of the minimum accretion rate for MSP formation}

It is noticed that only three MSPs, J0514-4002A, J1801-3210 and J1518+4904, lie below
the spin-up line with $\mdot=10^{15} g/s$. With the data from ATNF and Eq.(\ref{P}),
their positions correspond to the mean accretion rates
$0.6\times 10^{15}g/s, 0.8\times 10^{15}g/s$ and $0.9\times 10^{15}g/s$, respectively.
%It can be explain like this that the Hubble time exists error to be modified.
%
Possibly, %these three MSPs formed with accretion rates of $\mdot=10^{15} g/s$, and
their spin periods evolved to the present values due to the spin-down by of the electromagnetic
emission. This suggest that there exists a critical minimum accretion rate
$\mdot=10^{15} g/s$ for a MSP formation, and the arguments for this view are described
in the following.

At first, the real ages of MSPs have not yet been satisfactorily determined, although
their characteristic ages can be a little longer than the Hubble age \citep{cam94}.
During the MSP spin-up evolution, the minimum amount of accreted mass required for a MSP to
reach the spin period of several milliseconds is about $\Delta M_{cr} \sim 0.1 - 0.2 \ms $
\citep{wang11, tau12},
%\citep{wang11},
and the time of spin-up should be less than the Hubble time of about $t_{H} = 10^{10}$yrs
of a MSP. Thus, a critical minimum mass accretion rate for the MSP formation can be
estimated as:

\be \mdot_{cr} = \Delta M_{cr} / t_{H} \simeq 10^{15} g/s\;. \ee
If the accretion rate is much lower than this critical value, the NS will have no chance
to accrete sufficient mass to be spun-up to a period of a few milliseconds during
the evolution.  Based on
the above argument, a critical minimum accretion rate $\sim 10^{15} g/s$ is suggested as
a necessary condition for the MSP formation.

\section{Binary Pulsars in the B-P Diagram}
The evolution of the binary parameters of BPSRs, such as the companion mass, eccentricity
and orbital period, are topics of interest.  However, the BPSRs in GC can not
help us to understand their properties very well \citep{hes08, tau12}  since the stellar density
of the cluster is much higher than that of the Galactic disk {\bf which increase the collision} and capture
of the stars.
In this part, we ignore the 75 BPSRs in the GC, % and 1 BPSR in extra galaxy,
and study the distribution of 103 BPSRs in the Galaxy disk (GD) that have spin
period $P$ and magnetic field $B$ in the B-P diagram.

\begin{figure*}
\centering $\begin{array}{c@{\hspace{0.1in}}c@{\hspace{0.1in}}c}
\multicolumn{1}{l}{\mbox{}} & \multicolumn{1}{l}{\mbox{}} &
\multicolumn{1}{l}{\mbox{}} \\
\includegraphics[width=5cm]{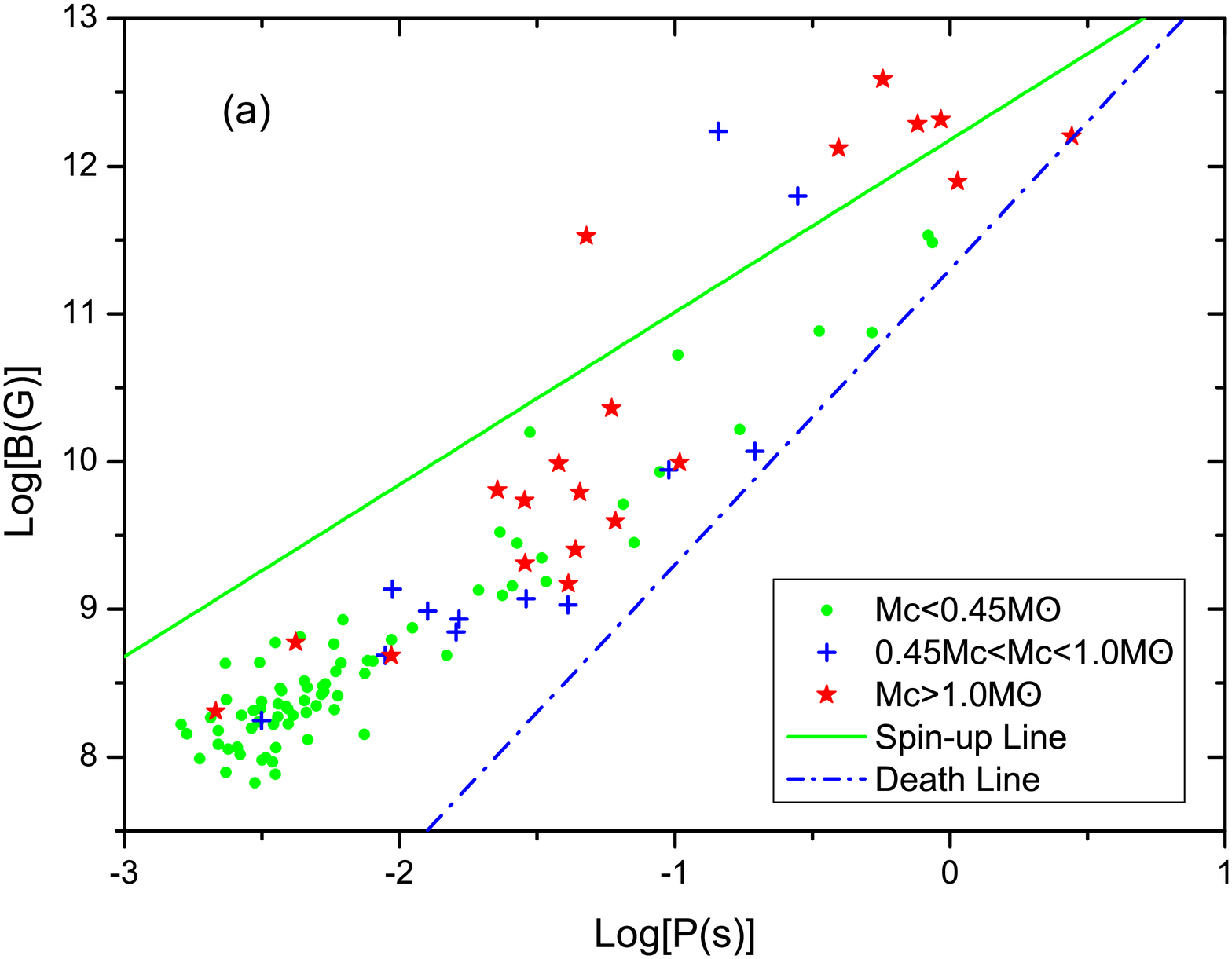} &\includegraphics[width=5cm]{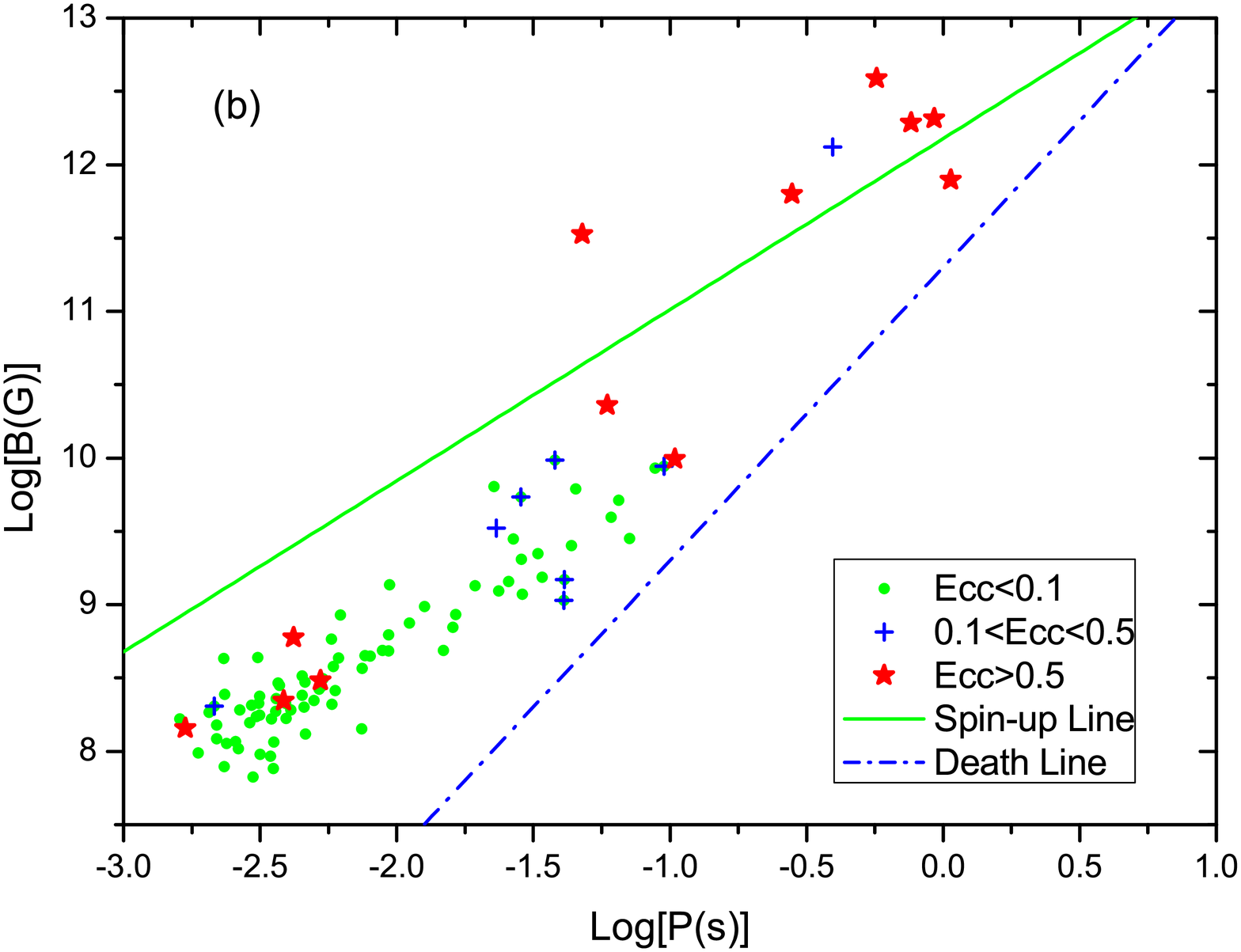}
&\includegraphics[width=5cm]{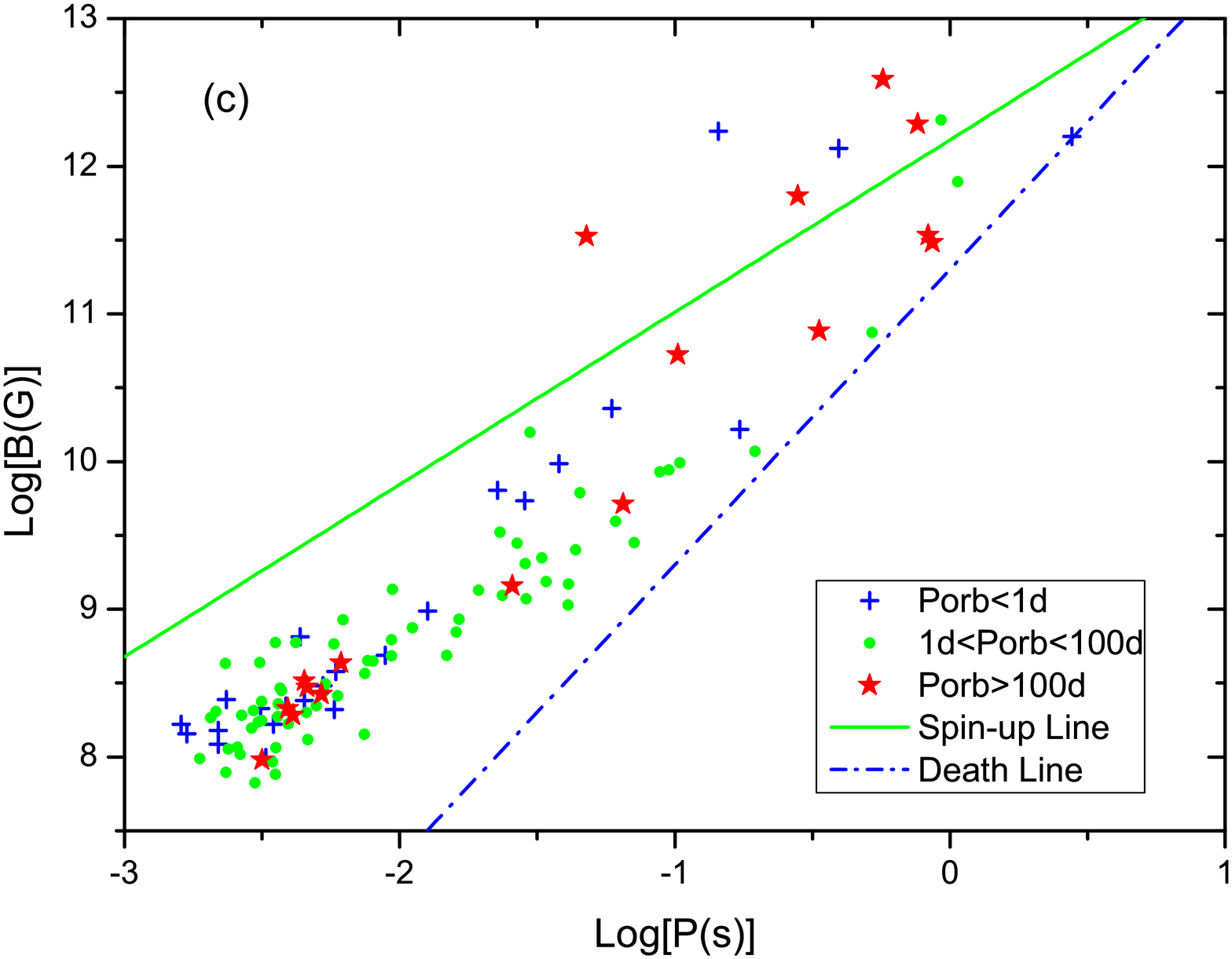}\\
\end{array}$
\caption{ B-P diagrams plotted with various parameter conditions of 103 BPSRs,
as shown in the figures, e.g. companion mass (a), eccentricity (b) and orbital period (c).
About $86.4\%$ of the BPSRs {\bf have low} magnetic fields $B<10^{10.5} G$,
which is considered as the threshold for the normal PSRs \citep{har97}.} \label{emo}
\end{figure*}

\begin{figure}
\includegraphics[width=7cm]{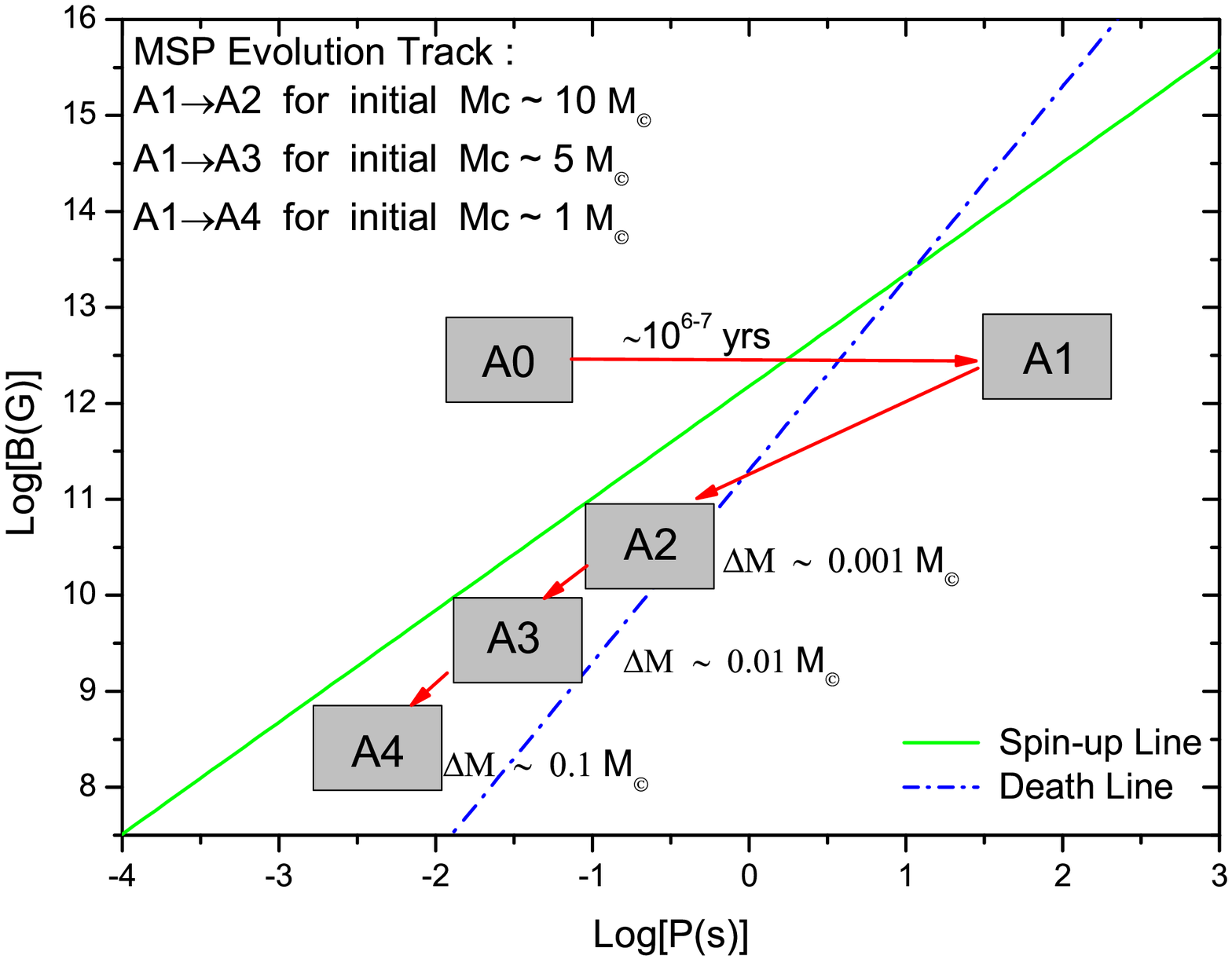}
\caption{Evolution tracks of binary pulsars in B-P
diagram.}\label{evolve}
\end{figure}

\subsection{BPSRs with the Different Parameter Conditions}

%\citep{har97}.
In Fig. \ref{emo}a, we notice that the BPSRs  with the small and
inter-mediate companion mass, $M_c < 0.1M_{\odot}$ and $0.1M_{\odot}<M_c < 0.45M_{\odot}$ as defined
by Stairs (2004) occupy about $12.6\%$ and $61.2\%$ in the total BPSRs, respectively.
%$77.6\%$ of the BPSRs with the companions less than $0.45M_{\odot}$ are MSPs.
In Fig.\ref{emo}b, the percentages of the BPSRs with the low eccentricity ($e<0.1$),
mediate eccentricity ($0.1<e<0.5$) and high eccentricity ($e>0.5$) occupy $85.1\%$, $6.9\%$
and $8.0\%$ of the total, respectively, which indicates that the eccentricity of BPSR has
deceased in the evolution process.
In Fig.\ref{emo}c, most of the BPSRs with the orbital period $P_{orb}<100 d$
share the spin period less than $\sim 100 ms$. Then those with long orbital
period $P_{orb}>100 d$ are distributed from $\sim30ms$ to $\sim1s$.
%In a binary system with a
%companion mass less (more) than the NS, the orbital period would be inclined to increasing
%(decreasing) in case the angular momentum of the binary system is conserved during the
%accretion process \citep{bha91}.
%The duration of the orbital period evolution of the binary pulsar with low-mass companions is
%mainly determined by the initial period \citep{bha91}.
%For the long initial
%period, the mass translate lasts only a few million years. Such that the magnetic field will
%not have time to decease. This is why the distribution of the BPSRs with long orbital period
%($P>100d$) is very diverse in Fig.\ref{emo}c.

\subsection{Evolution Scenario of Pulsar in Binary System}

There are many cases for the evolutions of binary systems \citep{lip84, tau11, tau12}.
The companion mass plays an important role in the evolution of binary system \citep{taa86, shi89}.
%\citep{shi89, taa86}.
%{\bf For most binary systems, the beginning of the evolution is always the case that}
%the two main sequence stars in the binary system start to evolve.
%and the heavier one evolves more quickly than the lighter one.
The star with a mass over about 8 solar masses experiences a supernova explosion firstly,
leaving a NS with a high magnetic field of $\sim 10^{12-13}G$ and spin period of $>10$
milliseconds, like the Crab pulsar ($B\sim 10^{12}G$, $P\sim 30 ms$). Such a new born pulsar
in the B-P diagram lies in the region A0 in Fig.\ref{evolve}, from where the pulsar
starts to evolve to A1 on account of the electromagnetic emission with a more or less constant
magnetic field within the time about $\sim 10^{7} yr$.

The life time of NS spin-up will be inversely related to the mass of its companion,
since heavier stars have shorter life spans,
which can be an approximated by the formula of the main sequence age
$T \sim 10^{10} (yr) (M/\ms)^{-2.5}$.
%
%After the first evolution stage from A0 to A1,
With the accretion mass provided by the companion, the NS will be spun up. The evolution can
be classified in the following three types \citep{shi89, taa86}.

(1) High Mass X-ray Binary (HMXB) case: the companion mass is about 10-25$\ms$.
The binary system will be the PSR+massive (PSR+NS) star if the system gets the accretion mass
$\sim 0.0001\ms-0.001\ms$ ($\sim 0.01\ms$),
while the evolution path might be A1-A2 (A1-A2-A3).
%The accretion material about $\sim 0.0001\ms-0.001\ms$ will make the NS spin up and
%evolve from A1 to A2 to form the NS+massive or NS+NS star system.}
The accretion evolution will stop at the spin-up line, where the stellar spin frequency
equals the orbital Keplerian frequency at the edge of the magnetosphere.

(2) Intermediate Mass X-ray Binary (IMXB) case: if the companion mass is in the
range of 3-10$\ms$, the NS spin-up evolution will continue the procedure of HMXB
case following up the route A1-A2-A3 or even to position A4 along the spin-up
line. The NS will accrete about $\sim 0.001\ms-0.1\ms$ to form
a NS+WD(heavier) systems, with the magnetic
field $B\sim 10^{9-10} G$ and the spin period $P\sim 20-100ms$.

(3) Low Mass X-ray Binary (LMXB) case: if the companion mass ranges at $\sim 0.8-3 \ms$, the NS
will evolve to the position A4 with the magnetic field $B\sim10^{8-9} G$ and the spin period $P<20ms$,
while the NS accretes $> \sim 0.1-0.2 \ms$ to form a MSP+WD(lighter) binary system
\citep{zhang11}.

From the above mentioned scenarios of BPSR evolutions, the companion mass and
accretion rate might play important roles in the evolution of binary
system \citep{wij97, shi89, taa86}.
%\citep{wij97, shi89, taa86}.

\section{Binary pulsars above he spin-up line}

\begin{figure}

\includegraphics[width=7cm]{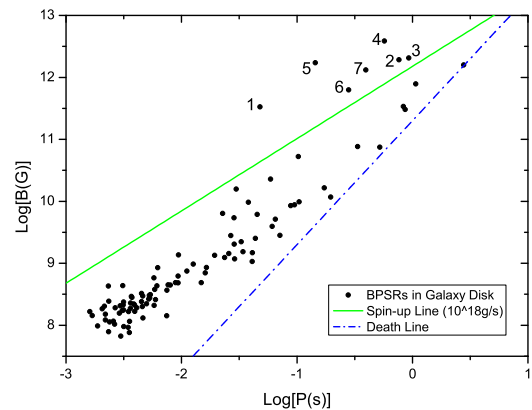}
\caption{B-P diagram for 103 BPSRs with the spin period and magnetic field
in GD. The plots share the similar meanings to those of Fig.1.}
%The two horizontal lines represent $B = 10^{10.5} G$ and $B = 10^{11.5} G$, respectively.}
\end{figure}\label{2}

In the binary systems, generally, the PSRs will evolve below the spin-up line after accreting a
sufficient amount of mass from their companions.
However, seven BPSRs lie above the Eddington spin-up line ($\mdot=10^{18} g/s$)
with the magnetic fields $>10^{11.5} G $, as shown in Fig.\ref{2}.
Their parameters are listed in Tab.~1.
According to their companion masses, these seven BPSRs are divided them into two types:
the ones with the massive stars ($M>4.0M_{\odot}$) and the others with the degenerate
stars (NS or WD).

\subsection{Type-I: BPSRs with Massive Companions}
Type-I includes four binary pulsars. Three of them (No.1,3 and 4 in Fig.\ref{2})
are with B or Be star companions
\citep{kas94, man01, sta01, sta03, wang04}.
PSR J1638-4725 (No.2 in  Fig.\ref{2}) has a massive star as its companion ($M_c>8.08M_{\odot}$)
which is in the main sequence stage. Its radio emission disappears near the periastron because
of the absorption  and scattering by the dense stellar environment. During the radio-quiet phase,
X-rays that are produced by the accretion matter onto the NS magnetosphere are likely to be
detectable \citep{mcl04}. By considering the parameters of this pulsar, such as the strong
magnetic field ($B\sim 1.93\times10^{12} G$),  not long
characteristic age ($\tau=2.53 Myr$),  very long orbital period
($P_{orb}=1940.9 d$) and very high eccentricity ($e=0.955$),  we
think that it has little possibility to accrete the matter from its
companion star, therefore it is inferred  that this pulsar
is a non-recycled BPSR,  born at  the position near A0.

The common characteristics of these four BPSRs are: they all have high
eccentricities, long orbital periods, massive companions in the main sequence stage. In all
cases the pulsars are new born ones with spin-down ones of less than a million years,; they
are not (yet) accreting, and are in the "non-recycled" phase. These four BPSRs are on the way
of evolution from A0 to A1 as shown in Fig.\ref{evolve}, where the first born PSRs are
experiencing the spin-down with little accretion mass.

\subsection{Type-II: BPSRs with  Degenerate Star Companions}

J1906+0746 (No.5 in Fig.\ref{2}) is a non-recycled PSR ($1.25M_{\odot}$)
in a double neutron star system. Its companion is a recycled PSR with the mass of
$1.37M_{\odot}$ \citep{lor08}. The short characteristic age
of J1906+0746 ($\tau=113000 yr$) indicates that it is a recently
formed young PSR after the core collapse, which should be the reason why its B-P position
lies above the spin-up line ($10^{18} g/s$).

B1820-11 (No.6 in Fig.\ref{2}) possesses  a slightly  massive
WD ($M_c=0.78 M_{\odot}$) \citep{por99, sta04} as its companion.
From its parameters ($\tau=3.22 Myr$, $B=6.29\times
10^{11} G$ and $P=279.829 ms$),
we suggest that it is a young recycled PSR.
The evolution history of this PSR can be understood in this way:
the initial magnetic field of NS can be as high as $B \sim 10^{13}
G$, and it evolves from the position A1 to A2 after accreting the
companion mass about $\sim0.001 \ms$, while one to two magnitude orders
of magnetic field has been deducted. Due to its high eccentricity and
large orbital period, B1820-11 might not get enough accreting mass from
its companion. In Eq.(\ref{P}) it is assumed that the spinning axis is parallel to the
magnetic axis, whereas in Eq.(\ref{Bsim}) the perpendicular condition is assumed.
Therefore the position of the spin-up line will be fuzzy because  the angle of the
spinning axis to the magnetic axis is unknown. The position of PSR B1820-11
might be below the spin-up line.

J1141-6545 (No.7 in Fig.\ref{2}) is a PSR of mass 1.3
\ms with an optical WD ($1.02\ms$) as its companion in the binary
system \citep{ant11}.
%\citep{ant11}.}
With the short orbital period ($P_{orbit}=0.1977 d$) and slightly low
eccentricity ($e=0.1719$), it can be derived that this PSR
acquired the accretion mass easily and followed up the recycled
process. Its position upon the spin-up line can be explained by
two possible reasons: the first one is the same reason
as for B1820-11, and the second one is that the PSR may have
experienced the accreting induced collapse (AIC) of a white dwarf
(WD) \citep{heu95, heu09}.

It is noted that the  spin-up line with Eddington rate in B-P diagram is also
influenced by the spin-up torque and radiation pressure in disk, which
 has been studied by some researchers \citep{kul08, gho92, lam05}.

\begin{table*}
 \centering
 \begin{minipage}{130mm}
  \caption{Parameters of Seven Binary Pulsars  above the Spin-up Line}
  \begin{tabular}{@{}lccccccccl@{}}
  \hline\hline
     No.& Name& $P/ms$ &$P_{orb}/d$   &e    &$M_C/M_{\odot}$        &$\tau/yr$ &$B/G $ &type\\

  \hline

   1& B1259-63& 47.763&         1236.724&   0.8699&     4.14&       332000& 3.34E11&NSMS\\
   2&J1638-4725&    763.933&    1940.9&     0.955&      8.08&       2.53E6& 1.93E12&NSMS\\
   3&J0045-7319&    926.276&    51.1695&    0.8079&     5.27&       3.29E6& 2.06E12&NSMS\\
   4&J1740-3052&    570.31&     231.0297&   0.5789&     15.82&      354000& 3.86E12&NSMS\\
   5&J1906+0746&    144.072&    0.166&      0.0853&     1.37&       113000& 1.73E12&DNS\\
   6&B1820-11&  279.829&        357.762&    0.7946&     0.78&       3.22E6& 6.29E11&NSWD\\
   7&J1141-6545&    393.899&    0.1977&     0.1719&     1.02&       1.45E6& 1.32E12&NSWD\\

\hline\hline
\end{tabular}
\end{minipage}
\end{table*}

\section{Discussions and Conclusions}

In this paper, we investigate the 186 BPSRs (136 MSPs) in the B-P diagram.
The origin and evolution information of their progenitors can be inferred by
analyzing their  positions relative to the spin-up lines in the B-P diagram.
The following results have been obtained:

(1) The observations of X-ray NSs in binary systems show that $\mdot=10^{17} g/s$
is  a conventionally detected mass accretion rate \citep{kli00, liu07}.
%(van der Klis 2000; Liu, Paradijs
%\& van den Heuvel 2007).
% \citep{kli00, liu07}.
%
In the B-P diagram, most BPSRs are    distributed around the spin-up
line for
$\mdot=10^{16} g/s$, %and above the limit spin-up line for $\mdot=10^{15} g/s$,
which means that most BPSRs should be born at the positions near the spin-up line
($\mdot=10^{17} g/s$), from where the PSRs experience the spin-down time of
$\sim10^{8-9} yr$ after the onset of the radio PSR emissions.
%
%
% (1) Most BPSRs
%distribute below the limit spin-up line ($\mdot=10^{18} g/s$) and
%above the minimum spin-up line ($\mdot=10^{15} g/s$), and mainly
%center around the spin-up line for $\mdot=10^{16} g/s$.
%
%Such a distribution seems to hint that
%

%
%Therefore, the derived accretion rates for the radio BPSRs are
%consistent with those observed in  X-ray NS binaries,  which
%supports the recycled theory for MSP formation.
%

(2) To form a MSP, the  mass of about at least $\sim$ 0.1-0.2 \ms  should be accreted
from the companion \citep{ wang11, zhang11}. Thus, within a Hubble time, the required
minimum accretion rate is $\mdot=10^{15} g/s$. In other words, if the accretion rate is
less than this critical value, a NS will  accrete  the mass less than 0.1 \ms and may
evolve to a recycled PSR with a slightly longer spin period than 20 ms. Furthermore, we
stress that this critical minimum accretion rate is an averaged value in the  whole binary
accretion history.

(3) It is noticed that the position of a pulsar and its spin-up line in the B-P diagram are
slightly affected by the choices of the NS mass and radius, and the radius has more
influence than the mass. However, unlike the measurements of the NS masses, the NS radius
has not yet been satisfactorily determined from observations \citep{zhang07, zhang11}.
%\citep{zhang07, zhang11}.
It is usually estimated to be in the range of 10-20 km, thus we cannot sure if the standard
value of $R=10 km$ introduces a big error in defining the PSR position in the
B-P diagram.

(4) The classification of the BPSRs by the conditions of the companion mass, eccentricity
and orbit period have given us a clear illustrations of their properties in B-P diagram.
Most millisecond BPSRs with small companion masses ($<0.45\ms$) should be long lived, such as
$10^{8-9} yr$, while $\sim0.1-0.2\ms$ is accreted during the accretion phase.
In addition, the majority millisecond BPSRs share the small eccentricities $e<0.01$, which
implies that their orbits were circularized during the recycle process. In fact, there have
already been many researches on this, e.g. by Tauris \& Sacinije (1999),
Tauris, van den Heuvel \& Savonije (2000), and Tauris, Langer \& Kromer (2012).
However, there is no clear and simple relation between the orbital period and BPSR evolution,
which hints that the evolution of the orbital period may be complicated in the recycle process.

(5) To interpret why the seven BPSRs  lie above the limit spin-up
line, we start with the binary histories of these sources. The first four
BPSRs have the heavy star companions ($M_c > 4 \ms$) staying at the main
sequence phases, and they are still spinning down via magnetic dipole radiation.
The recycle processes of these systems have not yet started, so they still go
on the evolutionary tracks of the first pulsar spin-down.
The fifth BPSR lies in a  double NS system  (NS+NS). However, unlike the system
PSR 1913+16 in which  the recycled pulsar has been observed, the observed pulsar
J1906+0746 should be the normal and the secondly formed pulsar.
Namely, its companion should be  a recycled NS that has not yet been
 observed as a pulsar.
Thus this PSR is experiencing the spin-down by the electromagnetic
radiation  towards the death line.
Both the sixth and seventh pulsars are similar, which have  the
degenerate white dwarfs  (NS+WD) as companions.
They might be the newly formed recycled  pulsars  with the Eddington
rate, while either the particular accretion torque and disk
structure or large NS radii may be responsible for their slightly
passing through the spin-up line.

In summary, the study of BPSRs in B-P diagram has provided us the
clues on their accreting evolution histories, which is a helpful
tool in constructing  a link between the radio MSPs and their
progenitors, where the influences by their binary parameters can be
properly investigated. The 186 binary pulsars from the 2008 pulsar
sample have been analyzed, and more samples are required to  present
robust conclusions on the evolutions of binary pulsars. With the
recent  construction of the powerful radio telescope, five hundred
meter aperture spherical telescope (FAST), more than 4,000 new
pulsars are expected to be observed in this decade  \citep{ nan06,
smi09, nan11}, while there will be a big sample set of binary
pulsars, e.g. about 500 ones, to show a clear trend of MSP
evolution.

\section*{Acknowledgments}
This work is supported by the National Basic Research Program of China (2012CB821800, 2009CB824800),
the National Natural Science Foundation of China (NSFC 11173034),  and CAS Knowledge
Innovation Project (KJCX2-YW-T09). We are grateful for the critic comments and advices from
van den Heuvel E. P. J., which greatly improves the quality of the paper.

\end{document}